# Sonochemical Modification of the Superconducting Properties of MgB$_2$

Tanya Prozorov[1,2], Ruslan Prozorov[#,1], Alexey Snezhko[1], Kenneth S. Suslick[2]

[1]*Department of Physics & Astronomy and USC NanoCenter, University of South Carolina, Columbia, SC 29208*

[2]*School of Chemical Sciences, University of Illinois at Urbana-Champaign, Urbana, IL 61801*

(Submitted May 2003)

Ultrasonic irradiation of magnesium diboride slurries in decalin produces material with significant inter-grain fusion. Sonication in the presence of Fe(CO)$_5$ produces magnetic Fe$_2$O$_3$ nanoparticles embedded in the MgB$_2$ bulk. The resulting superconductor-ferromagnet composite exhibits considerable enhancement of the magnetic hysteresis, which implies an increase of vortex pinning strength due to embedded magnetic nanoparticles.
[PACs: 74.81.Bd, 74.25.Ha, 74.25.Qt, 74.70.Ad]

Controlled modification of the pinning properties of bulk granular superconductors is an active area of applied and fundamental research[1-8]. Doping with different metals[9,10], variation of stoichiometry[11] and non-superconducting phase precipitation[12] are recent examples of the chemical tuning of superconducting materials. Systematic modification of superconductor *morphology* provides another way to influence inter-grain coupling and intra-grain critical currents[1,6,13].

Various techniques to control pinning properties of MgB$_2$ have been suggested.[14] Alternative synthetic routes[13,15] and post-synthesis treatments[16], fabrication of dense wires[17], pellets[16] and tapes[18], annealing in Mg vapor[19], doping with Na[20], Co, Fe[21], Cu, or Ag[10], introduction of SiC nanoparticles[22], Ag powder[23], Ti precipitates[24], synthesis of MgB$_2$/Mg nanocomposites[25], intra-layer carbon substitution[11,26] have all been reported.

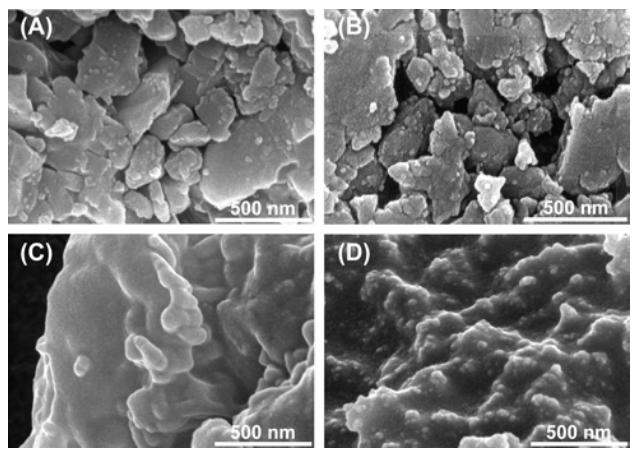

FIG. 1. Scanning electron images of: **(A)** original MgB$_2$ powder, sample A; **(B)** MgB$_2$ pellet, sample AP; **(C)** MgB$_2$ sonicated in decalin, sample S1; and **(D)** MgB$_2$ sonicated in decalin with Fe(CO)$_5$, sample SF1.

In this Letter, we report the sonochemical modification of grain morphology and intergrain coupling

---

[#]Corresponding author. e-mail: prozorov@mailaps.org

---

of polycrystalline MgB$_2$. The method is further extended for *in situ* synthesis and embedding of ferromagnetic nanoparticles, which are shown to act as efficient magnetic vortex pinning centers.

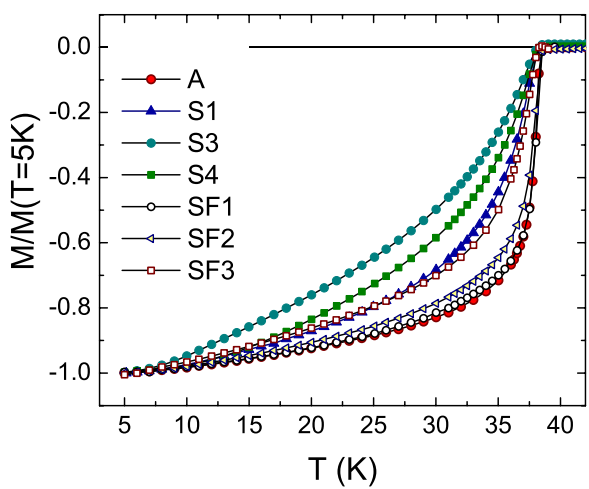

FIG. 2. Zero-field cooled magnetization measured in H = 10 Oe, normalized to its value at 5 K. The paramagnetic contribution for SF samples was subtracted using Curie-Weiss law measured up to 150 K.

In ultrasonically-irradiated slurries, turbulent flow and shock waves are produced by acoustic cavitation. The implosive collapse of bubbles during cavitation results in extremely high local temperatures (~5000 K)[27-28] and also creates high-velocity collisions between suspended particles with effective temperatures at the point of impact of ~3000 K.[29] These high velocity collisions cause localized inter-particle melting and "neck" formation[27-29]. The estimated speed of colliding particles approaches half of the speed of sound. MgB$_2$ polycrystalline powder (325 mesh, Alfa Aesar) was ultrasonically irradiated for 60 min at -5°C in 15 ml of decalin (0.13 %wt, 0.26 %wt, 0.5 %wt, and 2 %wt, respectively, at 20 kHz and ~50 W/cm$^2$) under ambient atmosphere using direct-immersion ultrasonic horn (Sonics VCX-750). A similar set of slurries was sonicated with the addition of 1.8 mmol of Fe(CO)$_5$. The resulting material was filtered, washed repeatedly with pentane, and air-dried overnight.

Magnetic measurements were conducted using *Quantum Design* Superconducting Quantum Interference Device (SQUID) MPMS magnetometer. For magnetic measurements, the powder was sintered at room temperature at a pressure of 2 GPa for 24 hours. The average sample mass was 10 mg. The magnetic moment was normalized using the initial slope, $dM/dH$, measured at 5 K after zero-field cooling. The slope is proportional to the fraction of the superconducting phase. For materials without magnetic nanoparticles, such normalization gives volume magnetization. For composites containing Fe$_2$O$_3$ nanoparticles, the normalization was done after subtraction of the paramagnetic contribution.

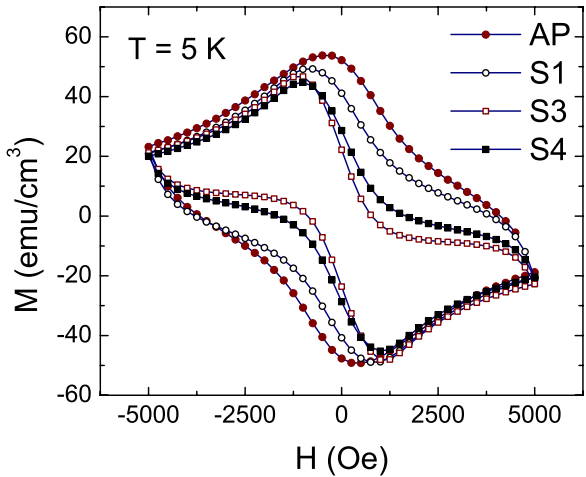

**FIG. 3**. Magnetization loops at *T*=5 K for sonicated samples S1, S2 and S3 compared to the original sample AP. Width of the hysteresis loops is reduced, but the Meissner expulsion is not.

Scanning electron micrographs (SEM) were taken on a Hitachi S-4700 instrument. Samples were additionally characterized by powder x-ray diffraction and differential thermal analysis. All reported results were reproduced on more than twenty-five samples. We use the following sample designations: original MgB$_2$ powder (A) and sintered pellet (AP); MgB$_2$ sonicated in decalin with various loadings of the slurry (pellets: S1, 0.13%wt; S2, 0.26%wt; S3, 0.5%wt; S4, 2%wt); MgB$_2$ sonicated in decalin with 1.8 mmol of Fe(CO)$_5$ (pellets SF1, SF2, and SF3 with the same loading of MgB$_2$ as S1, S2 and S3). SEM images of the original (A and AP) as well as sonicated samples (S1 and SF1) are shown in FIG. 1. Sample A is shown in FIG. 1(A) and sintered pellet AP, made of sample A, is shown in FIG. 1(B). No particular structural modification was observed. In contrast, sonicated powder used for sample S1, FIG. 1(C), and sonicated with Fe(CO)$_5$ sample SF1, FIG. 1(D), reveal distinctively modified morphology. Even though the decomposition temperature of MgB$_2$ (~1100 K)[14] is lower than the effective local temperatures achieved during transient cavitation, the initial material apparently undergoes surface melting, as implied by FIG. 1(D). This can be attributed to extremely high cooling rates (~10$^9$ K/s)[29,30] leading to formation of smooth welded grains in sonochemical process. In the case of a superconductor, such morphology change leads to better inter-grain coupling and annealing of the intra-grain defects, consistent with our observations. Sonication of MgB$_2$ powder in decalin with Fe(CO)$_5$ is accompanied by the *in situ* sonochemical formation of iron oxide nanoparticles[31] directly on MgB$_2$ grain surfaces, while concurrent ultrasound-driven melting results in embedding of Fe$_2$O$_3$ nanoparticles into the MgB$_2$ matrix, FIG. 1(D). The embedded particles act as efficient pinning centers where magnetic interaction with Abrikosov vortices provides extra force in addition to the core pinning. Similar enhancement was reported in 1966 for Hg-In alloys with mechanically dispersed Fe nanoparticles.[32,33] Related recent works study magnetic particles placed on the surface of low-$T_c$ superconducting films.[34-36] Our contribution is a novel way to embed ferromagnetic nanoparticles into high-$T_c$ materials.

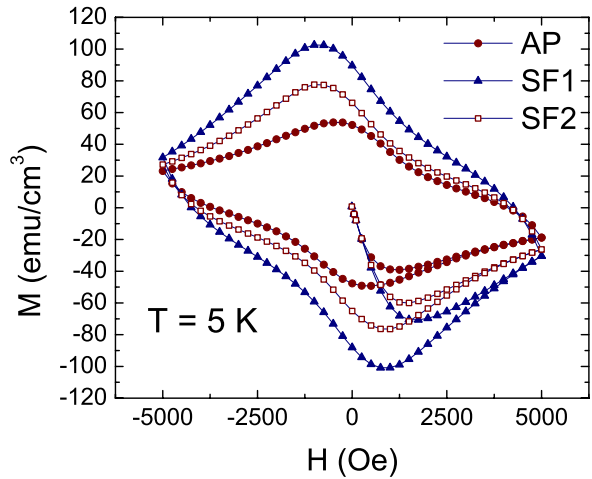

**FIG. 4.** Magnetization loops measured at *T*=5 K in MgB$_2$ sonicated in decalin with Fe(CO)$_5$. The hysteresis is largest for the lowest loading of MgB$_2$.

FIG. 2 shows *M(T)* curves measured in magnetic field of 10 Oe after zero-field cooling (ZFC). Superconducting transition temperature remains unchanged, $T_c \approx 38.5$ K. Curves in FIG. 2 are normalized by the magnetization value at 5 K and the paramagnetic contribution for the SF samples was subtracted. FIG. 3 shows the effect of sonication on the magnetization loops measured at 5 K for samples with different initial loading of MgB$_2$ slurries. The loops become less hysteretic and more asymmetric for loading up to 1% wt, after which the effect diminishes. This is what we expect for the material, where intra-grain defects are annealed during sonication and the large percentage of grains is fused together. This also provides the evidence that Meissner expulsion in granular superconductors is mostly due to intra-grain shielding and not weak inter-grain coupling.

As shown in FIG. 4, the situation is different for the samples sonicated with Fe(CO)$_5$. The magnetization loops are more hysteretic compared to sample A. However, the hysteresis decreases the increase of the MgB$_2$ loading. This is in agreement with the results of FIG. 3 where the optimum effect of sonication was achieved for 0.5% wt of MgB$_2$ slurry. FIG. 5 shows magnetization loops measured in sample SF3 at 30 and 42 K. The curve at 42 K is well described by the Langevin function, indicative of a superparamagnetic behavior. The hysteresis at $T=42$ K is due to some magnetic anisotropy and dipole-dipole interactions[37] of the dispersed Fe$_2$O$_3$ nanoparticles, and it is much smaller than the hysteresis due to pinning. We verified this conclusion by measuring remanent magnetization as a function of temperature.

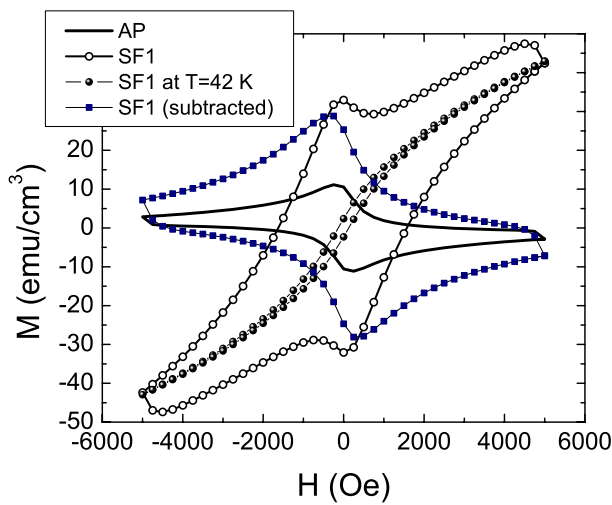

**FIG. 5.** <u>Open symbols</u> - measured magnetization loops for sample SF1; <u>filled squares</u> – same curve with paramagnetic contribution subtracted; <u>filled circles</u> – M(H) curve measured at T=42 K. The solid line is the M(H) curve of the unmodified MgB$_2$, sample AP measured at 30 K.

The irreversibility practically disappears at $T_c$. The difference, $\Delta M = M(30\,K) - M(42\,K)$, shown by solid squares in FIG. 5 is typical for a superconductor with significant pinning. The solid line shows magnetization curve of the original sample AP. The comparison indicates more than two-fold enhancement of pinning.

In conclusion, a novel method of a controlled modification of the superconducting properties of magnesium diboride is described. Ultrasonic cavitation leads to significant change in morphology without affecting chemical composition. Sonication in decalin results in granular superconducting material with significant inter-grain fusion and much less defective structure compared to the original MgB$_2$ powder. Sonication in decalin with the addition of Fe(CO)$_5$ produces a superconductor-ferromagnet composite in which ferromagnetic nanoparticles are embedded into the MgB$_2$ matrix. These particles act as efficient pinning centers. Our current research indicates that the described experimental technique and conclusions are applicable to other granular superconductors, such as YBa$_2$Cu$_3$O$_7$.

Discussions with V. Geshkenbein, B. Ivlev, E. Sonin and A. Koshelev are greatly appreciated. This work is supported by the NSF (EPSCoR Grant EPS-0296165 and CHE-0079124), a grant from the University of South Carolina Research and Productive Scholarship Fund, and the donors of the American Chemical Society Petroleum Research Fund. The SEM study was carried out in the Center for Microanalysis of Materials (UIUC), which is partially supported by the DOE under Grant DEFGO2-91-ER45439.